# The Cluster-in-Molecule Local Correlation Method with an Accurate Distant Pair Correction for Large Systems

Zhigang Ni,[1,2] Yang Guo,[3] Frank Neese,[4,5] Wei Li,[1] and Shuhua Li*,[1]

AUTHOR ADDRESS

[1] School of Chemistry and Chemical Engineering, Key Laboratory of Mesoscopic Chemistry of Ministry of Education, Institute of Theoretical and Computational Chemistry, Nanjing University, Nanjing 210023, China

[2] College of Materials, Chemistry and Chemical Engineering, Hangzhou Normal University, Hangzhou 311121, China

[3] Qingdao Institute for Theoretical and Computational Sciences, Shandong University, Qingdao 266237, China

[4] Max Planck Institut für Kohlenforschung, Kaiser-Wilhelm Platz 1, D-45470 Mülheim an der Ruhr, Germany

[5] FAccTs GmbH, Rolandstr. 67, 50677 Köln, Germany

AUTHOR INFORMATION

**Corresponding Author**

*Shuhua Li: Email: shuhua@nju.edu.cn

ABSTRACT

The cluster-in-molecule (CIM) local correlation approach with an accurate distant pair correlation energy correction is presented. For large systems, the inclusion of distant pair correlation energies is essential for the accurate predictions of absolute correlation energies and relative energies. Here we propose a simple and efficient scheme for evaluating the distant pair correlation energy correction. The corrections can be readily extracted from electron correlation calculations of clusters with almost no additional effort. Benchmark calculations show that the improved CIM approach can recover more than 99.97% of the conventional correlation energy. By combining the CIM approach with the domain based local pair natural orbital (DLPNO) local correlation approach, we have provided accurate binding energies at the CIM-DLPNO-CCSD(T) level for a test set consisting of eight weakly bound complexes ranging in size from 200 to 1027 atoms. With these results as the reference data, the accuracy and applicability of other electron correlation methods and a few density functional methods for large systems have been assessed.

**TOC GRAPHICS**

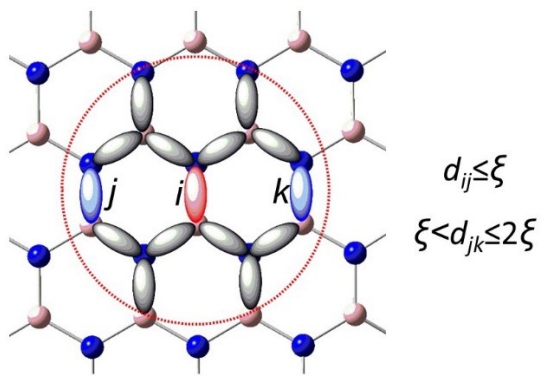

$$\Delta E^{\text{CIM}} = \sum_i \Delta E_i + \sum_{jk} \Delta \tilde{E}_{jk}$$

In computational chemistry, electron correlation plays an essential role in accurately describing various properties of molecules. However, the application of conventional electron correlation methods is limited to small and medium-sized systems due to their steep computational scaling with respect to the system size. For example, the commonly used second-order Møller−Plesset perturbation theory (MP2)[1] and the coupled cluster theory with singles, doubles, and perturbative triples excitation (CCSD(T))[2] scale as $O(N^5)$ and $O(N^7)$, respectively, where $N$ is a measure of the molecular size. Their computational costs can be significantly reduced by using localized molecular orbitals (LMOs) instead of canonical molecular orbitals (CMOs) with negligible loss of accuracy.

Various LMO based linear or low scaling correlation algorithms have been developed, which are collectively referred to as "local correlation methods".[3] Broadly speaking, there are two large families of local correlation methods that might be referred to as "direct" and "cluster-based" local correlation methods. In the first group,[3-24] the formalism and equations of the parent canonical methods are retained and the system is treated as a whole. The reduction in computational effort arises from carefully constructing the computational entities (LMOs, integrals, amplitudes, …) such that negligible contributions can be identified and their computation is avoided. Thus, in the 'direct' local correlation methods, a set of equations for the amplitudes of strongly interacting pairs have to be solved simultaneously. In the other family of local correlation methods,[25-51] the correlation energy of each LMO is computed within a small cluster, which consists of a subset of occupied and virtual LMOs distributing around this LMO. In principle, the correlation energy of the central LMO obtained from the cluster is a good approximation to the exact one, provided that the cluster is large enough. Compared to the 'direct' local correlation methods, cluster-based methods require much less memory and disk storages, and are easier to parallelize, because clusters can be constructed and treated independently. However, in the cluster-based methods, there are redundant calculations due to the overlap of clusters. It is worth noting that cluster-based calculations rely on the possibility that electron correlation calculation of each cluster can be done with the parent method.

The cluster-in-molecule (CIM) approach, as a representative cluster-based local correlation method, has been developed by Li and co-workers,[33-42, 52, 53] and other groups.[43, 54-56] In the CIM approach, the total correlation energy of the full system is expressed as the summation of correlation energies of all occupied LMOs,

$$\Delta E = \sum_i \Delta E_i, \tag{1}$$

where $i$ refers to an occupied LMO. To obtain the correlation energy contribution of the $i$th LMO (called as the central orbital), its spatially neighboring (within a distance threshold $\xi$) occupied orbitals (called as environmental orbitals) are included in the cluster to form the occupied space. Recently, we have proposed an efficient algorithm for the construction of the virtual space of a cluster.[42] Then, the occupied and virtual orbitals in a cluster are truncated from the full atomic orbital (AO) space to a smaller AO domain by using the Boughton−Pulay projection method.[57] After the cluster construction step, electron correlation calculations can be performed at any level within clusters to get the correlation energy of each central orbital. Taking the CIM-MP2 method as an example, within a cluster [$P$], the correlation energy contribution of LMO $i$ can be written as

$$\Delta E_i = \sum_{j \in [P]} \Delta E_{ij}. \tag{2}$$

where $i$ is the central orbital in cluster [$P$]. The pair correlation energy reads,

$$\Delta E_{ij} = \sum_{ab \in [P]} \left[ 2(ia|jb) - (ib|ja) \right] \tau_{ij}^{ab}, \tag{3}$$

where $j$ denotes environmental occupied orbitals, $a$ and $b$ virtual orbitals, $(ia|jb)$ the two-electron repulsion integrals, and $\tau_{ij}^{ab}$ excitation amplitudes. At CC levels, one can reformulate the standard correlation energy expressions as well. The detailed expressions can be found in our previous works.[41, 42, 55]

In the CIM approach, there is only one parameter, $\xi$, which controls the accuracy by controlling the size of clusters. In recent CIM applications, 5.5 Å has been chosen as the default value for a compromise between computational cost and accuracy. For three dimensional systems, with the default $\xi$, and a triple-$\zeta$ basis set, the largest cluster may have about 3000 basis functions. Although it is possible to use RI-MP2 to treat such clusters, it prohibits the calculation at the CCSD or CCSD(T) level. To overcome this difficulty, one has to further compress the occupied and virtual spaces of clusters before the CCSD(T) calculations of clusters. One method to reduce the MO spaces is the local natural orbital (LNO) method developed by Kállay and co-workers,[43, 45-47] in which the number of occupied or virtual MOs in clusters is reduced using natural orbitals. With an efficient implementation in the MRCC suite,[58] an LNO-CCSD(T) calculation with 44035 basis functions has been reported.[46] Alternatively, some of us have recently combined the CIM framework with the domain based local pair natural orbital (DLPNO) approach, which is a very popular "direct" local correlation approach, to solve this problem.[18-24, 55] In the CIM-DLPNO approach, the CIM scheme is used to partition the system into clusters and the DLPNO approach is employed to compute the correlation energies of clusters. The CIM-DLPNO-CCSD(T) method has been applied to a system with more than 20000 basis functions.[55]

In the CIM framework, a certain central LMO $i$ is only correlated with its spatially neighboring LMOs within distance $\xi$. Only the correlation energies between occupied orbital pairs within distance $\xi$ are taken into account in previous CIM implementations. If a pair of occupied LMOs separated by a distance beyond $\xi$ is defined as a distant pair, the correlation energy contribution from these distant pairs is completely neglected. Although the correlation energy from a distant pair is very small, the number of distant pairs increases quadratically with the number of LMOs.[59] Thus, for systems with hundreds or thousands of atoms, the correlation energy from all distant pairs may contribute significantly to the absolute correlation energy. The neglected contributions may add up to a significant energy that may not always cancel out upon taking chemically relevant energy differences. For example, when computing binding energies of a supermolecule containing several individual molecules, the complex and its components must be described with equal accuracy. Hence, the inclusion of distant pair correlation energies is critical in predicting accurate binding energies of large complexes.

There are usually two ways to recover the correlation energies of distant pairs. The first way is to evaluate the distant pair correlation energies accurately by including two-body interactions.[37, 51] And the second is to evaluate distant pair correlation energies approximately, for example, with the multipole moment expansion (MME) pioneered by Hetzer, Pulay, and Werner.[60] The MME approximation for distant pairs have been widely employed in a number of local correlation methods, like local MP2,[6] DLPNO algorithms,[22] LNO,[45, 46] as well as CIM-DLPNO method[55] implemented in ORCA. It was pointed out by Werner[59] that the MME approximation usually gives about half of the true correlation energy, and could cause large absolute errors for extremely large systems. Recently, the accuracy and efficiency of the MME approximation have been improved by introducing PNO approximations.[59]

In the present work, we report a new algorithm for estimating the distant pair correlation energies within the CIM framework. An illustrative cluster, in which $i$ is the central orbital and $j$, $k$, *etc.* are environmental orbitals, is shown in Fig. 1. With the distance threshold $\xi$, the diameter of the cluster is 2$\xi$. Once the electron correlation equations are solved within a given cluster, the correlation energies of all occupied orbital pairs can be obtained via Eq. (3). In the previous CIM implementations, the correlation energies between two environmental orbital pairs were never used. In the present work, the correlation energies of environmental orbital pairs with distance between $\xi$ and 2$\xi$ are extracted and taken as the distant pair correction. Thus, the improved CIM correlation energy can be expressed as

$$\Delta E = \sum_i \Delta E_i + \sum_{jk} \Delta \tilde{E}_{jk} \quad \left(\xi < d_{jk} \leq 2\xi\right), \tag{4}$$

with the distant pair correlation energies

$$\Delta \tilde{E}_{jk} = \frac{1}{M} \sum_P \Delta E_{jk}(P), \tag{5}$$

where $jk$ refers to an occupied LMO pair in cluster [$P$] with the distance between $\xi$ and 2$\xi$. It should be noted that a certain pair $jk$ may appear simultaneously in several clusters. However, since it is embedded in in different environments, the pair energies $\Delta E_{jk}$ will be slightly different from one cluster to the next. Hence, in eq (5), $M$ is the number of times that the $jk$ pair appears in different clusters and the mean value is used as the final pair correlation energy.

In the present CIM algorithm, distant pair correlations beyond 2$\xi$ are still neglected, which appears reasonable given that the pair interaction energies decay as $R^{-6}$. Importantly, the most important distant pairs are calculated at the same level as the central orbitals, which should be much more accurate than the MME method adopted by other approaches.[7, 8, 23, 24] This is a pressing problem in the study of weak intermolecular interactions, in particular since MP2 is known to not be accurate enough to provide quantitatively accurate binding energies for π-π interactions. Hence, MP2 can not be used to fully quantitatively correct local coupled cluster energies. We hope that this improved CIM algorithm can provide not only more accurate total correlation energies, but also more accurate relative energies or binding energies at various correlation levels for large systems.

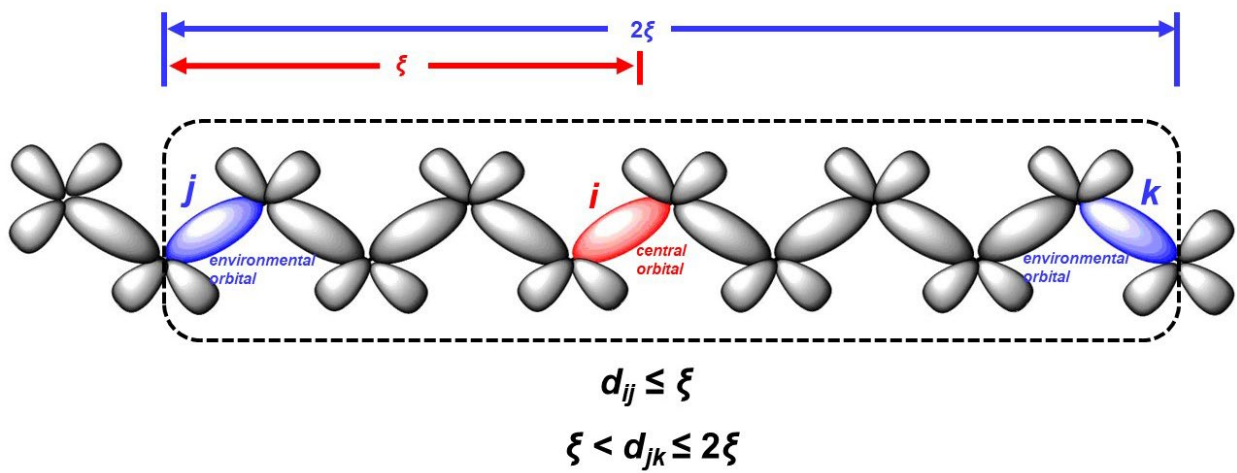

Figure 1. An illustrative orbital cluster (within the dashed line area) with *i* being the central orbital and *j* and *k* being two examples of environmental orbitals. In this cluster, the distance between *i* and any other occupied orbital is smaller than *ξ*. However, there are occupied orbital pairs, like *jk*, whose distances are between *ξ* and 2*ξ*.

The new CIM algorithm with this distant pair correction has been implemented in the LSQC 2.4 program,[61] at MP2 level. Both the resolution of identity (RI) MP2[62] and the spin-component-scaled (SCS)[63] MP2 have been implemented within the CIM framework (named as CIM-RI-MP2, CIM-SCS-RI-MP2). The new CIM-DLPNO-CCSD(T) algorithm is realized by interfacing LSQC to ORCA, invoking the DLPNO-CCSD(T) algorithm for cluster calculations. To obtain more accurate total energies, the DLPNO-CCSD(T) method with iterative (T) algorithm is employed.[24]

The accuracy of the newly implemented CIM-RI-MP2, CIM-SCS-RI-MP2, and CIM-DLPNO-CCSD(T) approaches was benchmarked against the corresponding parent methods (RI-MP2, SCS-RI-MP2, and DLPNO-CCSD(T)), respectively, using a benchmark set containing five medium-sized complexes formed by noncovalent interactions. The structures of these five complexes are shown in Fig. 2 (see Supporting Information (SI) for detailed descriptions of the structures). Throughout the Letter, the default distance threshold (5.5 Å) is still used in all CIM calculations and the TightPNO settings are employed in the (CIM-)DLPNO-CCSD(T) calculations. The absolute correlation energies are listed in Table S1 in SI. The percentages of the correlation energies from CIM approaches with respect to the corresponding parent methods are also provided. For all tested systems, the present CIM method recovers more than 99.97% of the absolute correlation energies. The percentages of the correlation energies recovered from the distant pairs with respect to the total correlation energies of the parent methods are listed in Table S2. It shows that the correlation energies of the distant pairs have significant contributions to the total correlation energies. The maximum percentage is 0.088% for CIM-RI-MP2 of system (e).

The wall clock times for the five complexes computed by CIM-RI-MP2 and CIM-DLPNO-CCSD(T) methods are compared with those of the corresponding parent methods in Table S3. In comparison with the RI-MP2 method, the CIM-RI-MP2 approach is computationally more efficient only when the system is large enough. In our testsuite, CIM-RI-MP2 is faster than the RI-MP2 for systems (a), (b), and (d). However, since DLPNO-CCSD(T) itself is a linear-scaling method, CIM-DLPNO-CCSD(T) is about five times (or more) slower than DLPNO-CCSD(T), due to the fact that many overlapping clusters should be calculated. Nevertheless, as there are still strong reasons to pursue CIM-DLPNO-CCSD(T). First, the memory and disk consumption of CIM-DLPNO-CCSD(T) is much lower than that of DLPNO-CCSD(T) such that much larger calculations can be done on a given hardware. Second, CIM-DLPNO-CCSD(T) will show much better parallel scaling on highly parallel platforms than DLPNO-CCD(T) itself can deliver. Thus,

CIM-DLPNO-CCSD(T) will be the method of choice for very large systems that cannot be treated by DLPNO-CCSD(T).

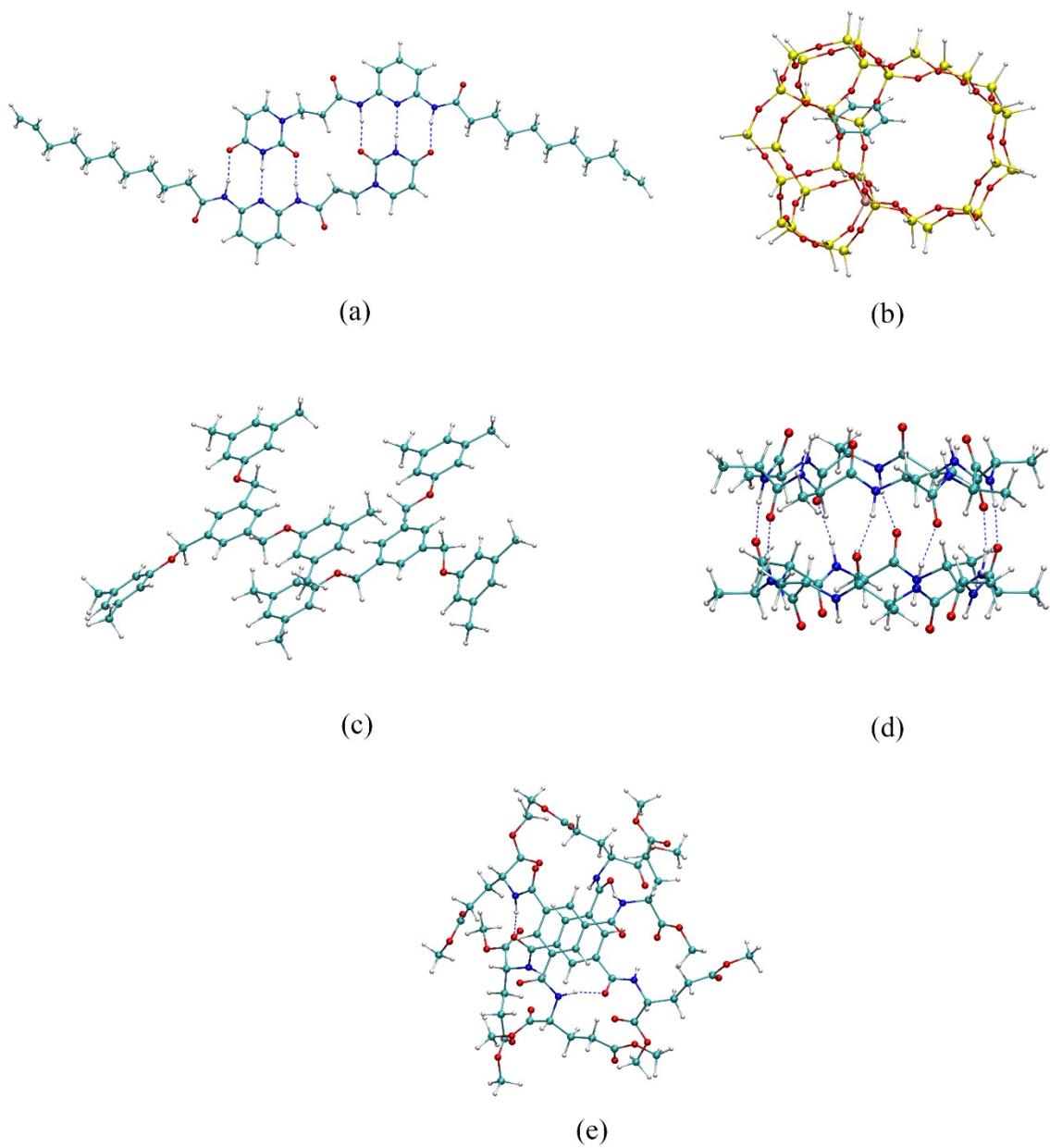

Figure 2. Five complexes in benchmark set.

The binding energies calculated by CIM-based methods are compared with those from the corresponding parent methods as well. Here, the binding energy is defined as

$$E_{\text{binding}} = E(\text{AB}) - E(\text{A}) - E(\text{B}), \quad (6)$$

where the structures of A and B are directly taken from the complex AB without relaxations. The extrapolated complete basis set (CBS) limit results are reported to compensate the lack of diffuse basis functions (See SI for details about the CBS extrapolation). The basis set superposition error (BSSE) should be small and is omitted here.

The binding energies are listed in Table 1 and the accuracy of binding energies obtained with various CIM methods are depicted in Fig. 3. It shows that the binding energies from the CIM approach are in excellent agreement with those from the full system results, with the largest deviation less than 2 kcal/mol. One may note that CIM-DLPNO-CCSD(T) has smaller errors than the other two methods. This is probably due to the fact that the DLPNO-CCSD(T) result is an excellent approximation to the full system CCSD(T) result (not available for such systems).

Table 1. Binding energies (kcal/mol) of complexes in the benchmark set computed by different levels of theory.

| System | CIM-RI-MP2/CBS | RI-MP2/CBS | CIM-SCS-RI-MP2/CBS | SCS-RI-MP2/CBS | CIM-DLPNO-CCSD(T)/CBS | CIM-DLPNO-CCSD(T)‖RI-MP2 | DLPNO-CCSD(T)/CBS |
|---|---|---|---|---|---|---|---|
| a | −44.48 | −44.91 | −38.15 | −38.50 | −43.04 | −42.65 | −43.17 |
| b | −14.71 | −15.63 | −7.97 | −8.78 | −11.77 | −12.91 (−12.94[a]) | −12.07 |
| c | −41.91 | −43.13 | −28.48 | −29.53 | −29.79 | −31.72 | −29.82 |
| d | −80.61 | −82.36 | −67.81 | −69.33 | −77.45 | −77.29 | −78.22 |
| e | −83.38 | −84.81 | −64.08 | −65.40 | −74.59 | −75.28 | −74.16 |

[a] The binding energy is calculated by using Eq (9) as the ΔCCSD(T) correction.

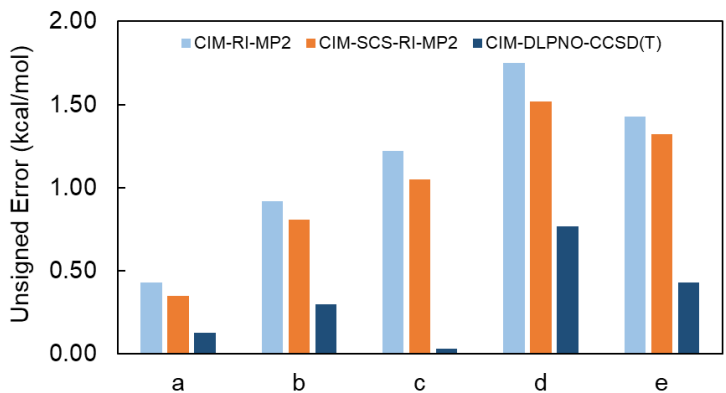

Figure 3. Unsigned errors (kcal/mol) of binding energies calculated by the CIM approach compared with the corresponding parent method.

We have also compared the binding energies calculated by the CIM approaches with and without the distant pair correction (absolute energies are listed in Table S4 in the SI). In Fig. 4, it can be seen that without distant pair correction, the deviations (relative to the full system results) increase significantly with the system size. However, with the distant pair correction, the deviations are reduced significantly. For example, the binding energy of the largest complex (e) calculated with CIM-DLPNO-CCSD(T) without distant pair correction deviates from the full system result by about 6.4 kcal/mol. After the distant pair correction is included, the deviation is reduced to 0.4 kcal/mol.

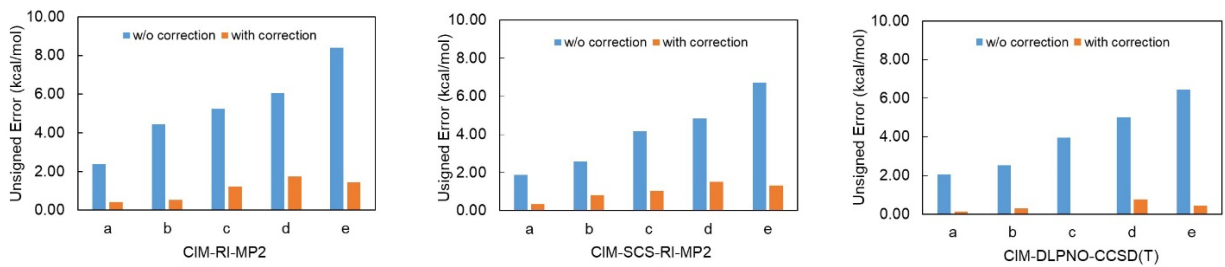

Figure 4. Deviations (kcal/mol) of the binding energies calculated by the CIM approaches with and without distant pair corrections at different levels (relative to the full system results).

For calculations with larger basis sets (e.g. cc-pVTZ and higher), DLPNO-CCSD(T) calculations for clusters are time-consuming. To further reduce the computational costs, we can approximately obtain the CCSD(T) energy by[64]

$$E_{\text{CIM-DLPNO-CCSD(T)}\|\text{RI-MP2}} = E_{\text{CIM(5.5)-RI-MP2/CBS}} + \Delta\text{CCSD(T)}, \quad (7)$$

where $\Delta$CCSD(T) is the CCSD(T) correction energy correction calculated with a moderate basis set, like cc-pVDZ. We have proposed two ways to obtain the $\Delta$CCSD(T) correction. The first way is to obtain the correction from CIM-DLPNO-CCSD(T) calculation of the full system

$$\Delta\text{CCSD(T)} = E^{\text{full}}_{\text{CIM(5.5)-DLPNO-CCSD(T)/cc-pVDZ}} - E^{\text{full}}_{\text{CIM(5.5)-RI-MP2/cc-pVDZ}}. \quad (8)$$

The second way is to further reduce the computational costs of Eq. (8), in which only the "active" site of the full system (such as catalytic centers and adsorption sites) is computed at CIM(5.5)-DLPNO-CCSD(T)/cc-pVDZ level. With this approximation, the $\Delta$CCSD(T) correction can be expressed as

$$\Delta\text{CCSD(T)} = E^{\text{active}}_{\text{CIM(5.5)-DLPNO-CCSD(T)/cc-pVDZ}} - E^{\text{active}}_{\text{CIM(5.5)-RI-MP2/cc-pVDZ}}. \quad (9)$$

This is equivalent to our previously proposed two-level CIM method,[35, 42] which has shown great success in studying reaction barriers.

The binding energies calculated by CIM-DLPNO-CCSD(T)‖RI-MP2 according to Eq. (8) are compared with respect to the DLPNO-CCSD(T) results in Fig. 5. For five complexes, the deviations are all within 2 kcal/mol. We have also computed the binding energy using Eq. (9) for system (b), in which the active region is shown in the SI. The result is −12.94 kcal/mol, being very close to the binding energy calculated with Eq. (8), −12.91 kcal/mol.

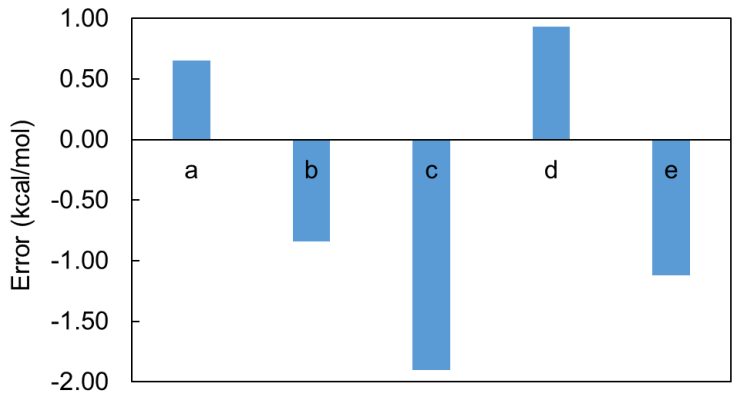

Figure 5. Errors of binding energies (kcal/mol) calculated by CIM-DLPNO-CCSD(T)‖RI-MP2 with respect to the DLPNO-CCSD(T)/CBS results.

Benchmark calculations show that the new CIM approach can achieve high accuracy for binding energy calculations by including the contribution from distant pairs. To further test the performance of this improved CIM algorithm, here we propose a test set, Extra-Large 8 (referred to as ExL8), consisting of eight large weakly bonded complexes with atoms ranging from 200 to 1027 (Fig. 6). There are mainly two types of non-covalent interactions in the eight complexes: $\sigma$-$\sigma$ dispersion and hydrogen bonding. Systems (1), (2), (5) and (6) are four supramolecular systems, among which the former three are dominated by hydrogen-bond interactions and system (6) is dominated by $\sigma$-$\sigma$ dispersion. Systems (3) and (4) are two cluster models of absorption complexes. Systems (7) and (8) are two typical biomolecular systems, DNA double helix and protein-ligand complex. Detailed structural information about ExL8 can be found from the SI. The binding energies of systems in the ExL8 set computed by various CIM methods are listed in Table 2.

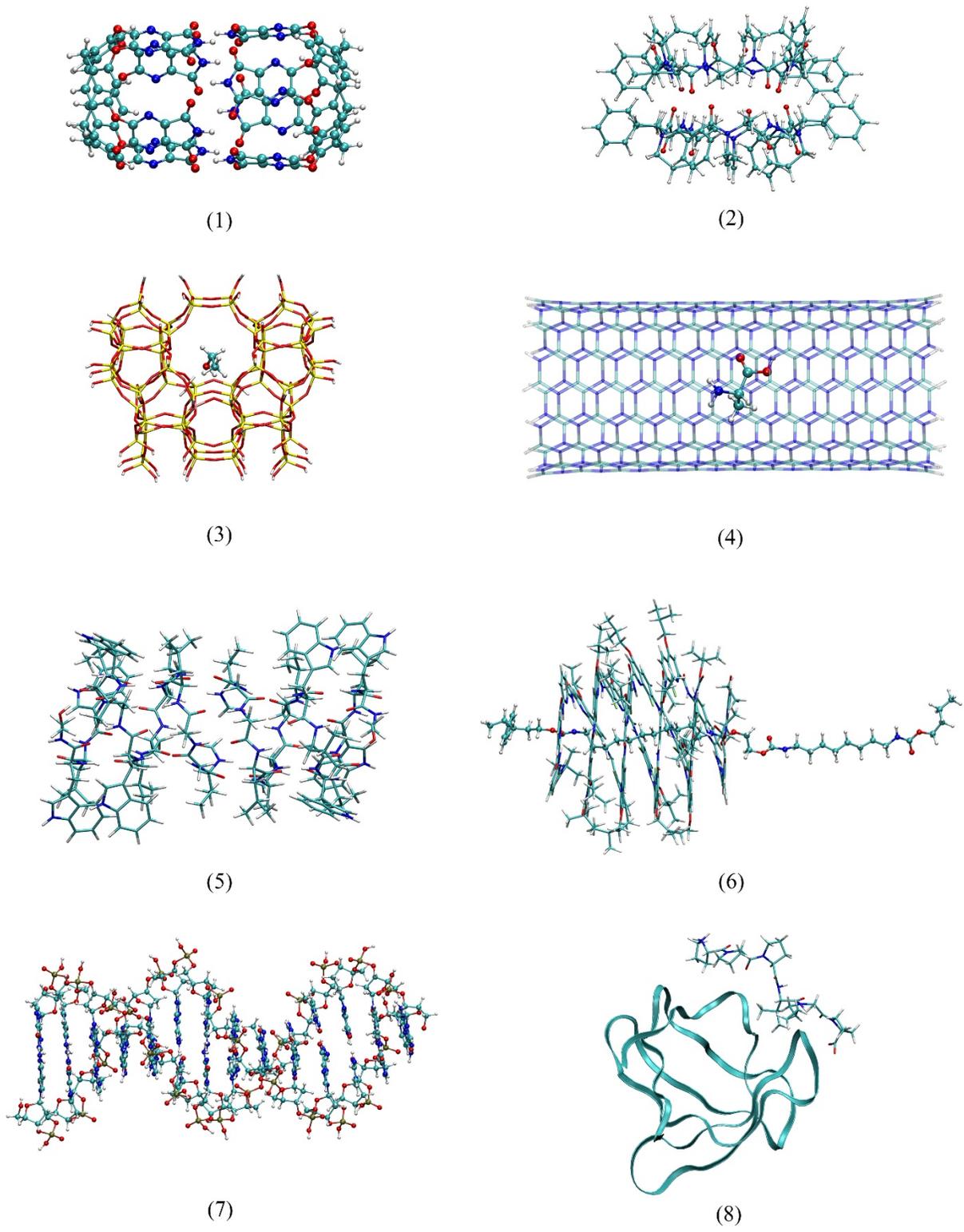

Figure 6. Complexes in ExL8 test set.

Table 2. Binding energies (kcal/mol) of the complexes in the ExL8 at various theory levels. The number of atoms ($N_{atom}$) and the number of basis functions ($N_{bas}$) in cc-pVTZ calculations are also listed.

| System | $N_{atom}$ | $N_{bas}$ | CIM-RI-MP2/CBS | CIM-SCS-RI-MP2/CBS | CIM-DLPNO-CCSD(T)‖RI-MP2 |
|---|---|---|---|---|---|
| 1 | 200 | 5360 | −69.94 | −59.56 | −65.67 [a] |
| 2 | 296 | 6576 | −75.68 | −58.10 | −69.63 [a] |
| 3 | 328 | 9072 | −37.77 | −32.97 | −36.55 [b] |
| 4 | 381 | 10806 | −19.53 | −14.69 | −17.83 [b] |
| 5 | 552 | 12080 | −43.09 | −34.02 | −40.13 [a] |
| 6 | 750 | 17316 | −88.50 | −66.18 | −78.80 [a] |
| 7 | 910 | 21932 | −422.60 | −380.19 | −416.08 [a] |
| 8 | 1027 | 22778 | −39.42 | −26.43 | −35.70 [b] |

[a] Calculated with Eq (8), where the ΔCCSD(T) energy is calculated from the full system.

[b] Calculated with Eq (9), where the ΔCCSD(T) energy is calculated from the active region. Atoms in active region are shown in the SI.

Table 3. Comparison of the binding energies (kcal/mol) of the complexes in the benchmark set and ExL8 computed with four DFT methods at the cc-pVTZ basis set and the reference method discussed in the text.

| System | B3LYP-D3 | ωB97X-D | M06-2X | M06-2X-D3 | Reference |
|---|---|---|---|---|---|
| a | −46.44 | −44.76 | −40.66 | −43.20 | −43.04 |
| b | −19.87 | −14.91 | −12.09 | −15.79 | −11.77 |
| c | −29.30 | −31.34 | −25.70 | −31.64 | −29.79 |
| d | −83.00 | −82.10 | −76.34 | −83.53 | −77.45 |
| e | −80.60 | −78.67 | −72.61 | −82.67 | −74.59 |
| 1 | −74.31 | −69.33 | −72.38 | −76.33 | −65.67 |
| 2 | −76.69 | −78.41 | −64.69 | −79.42 | −69.63 |
| 3 | −43.82 | −41.18 | −36.02 | −39.44 | −36.55 |
| 4 | −21.74 | −23.85 | −14.76 | −20.12 | −17.83 |
| 5 | −45.43 | −44.03 | −36.16 | −44.87 | −40.13 |
| 6 | −82.81 | −79.78 | −59.61 | −84.06 | −78.80 |
| 7 | −423.30 | −411.77 | −354.88 | −392.21 | −416.08 |
| 8 | −39.38 | −40.38 | −23.50 | −36.72 | −35.70 |

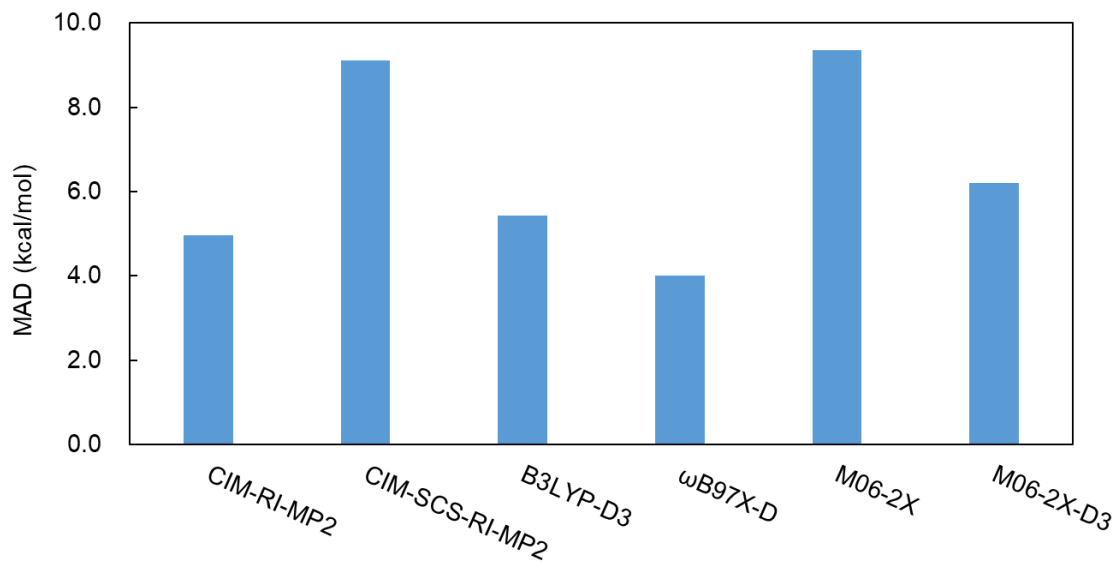

Figure 7. The MADs (kcal/mol) of the binding energies calculated by different methods.

With the CIM-DLPNO-CCSD(T)||RI-MP2 results for the ExL8 test set and CIM-DLPNO-CCSD(T)/CBS results for five complexes in the benchmark set as the reference data, we can now evaluate the performance of CIM-RI-MP2 and CIM-SCS-RI-MP2 in predicting the binding energies of medium-sized and large molecules. The mean absolute deviation (MAD) values are shown in Fig. 7. It is clearly seen that the CIM-RI-MP2 method has a much better performance than the CIM-SCS-RI-MP2 method, since its MAD (5.0 kcal/mol) is smaller than that (9.1 kcal/mol) for CIM-SCS-RI-MP2. However, we notice that CIM-RI-MP2 tends to overestimate the binding energies to some extent. For system (c) with π-π stacking interaction, the overestimation of CIM-RI-MP2 is quite significant, being about 12 kcal/mol. In contrast, the CIM-SCS-RI-MP2 method underestimates the binding energies. For the DNA double helix (system (7)), the binding energy from CIM-SCS-RI-MP2 is about 36 kcal/mol less than the CIM-DLPNO-CCSD(T)||RI-MP2 result.

For eight complexes in the ExL8 test set and five complexes in the benchmark set, we have also calculated the binding energies with four popular density functional methods which can provide reasonable description of non-covalent interactions. These functionals include B3LYP[65, 66] with Grimme's D3 correction[67] utilizing Becke−Johnson damping[68] (B3LYP-D3), ωB97X-D,[69] M06-2X,[70] and M06-2X with D3 correction using zero damping (M06-2X-D3). All DFT calculations are performed using Gaussian 16 program.[71] The binding energies computed at the cc-pVTZ basis set without BSSE correction are collected in Table 3, together with the reference data described above. The MAD values of these four DFT methods are also listed in Fig. 7. One can see that ωB97X-D provides the most accurate binding energies for the systems under study, with a MAD of only about 4.0 kcal/mol, and B3LYP-D3 has the second-best performance. In contrast, M06-2X has the largest MAD (about 9.4 kcal/mol) and it tends to underestimate the binding energies for systems with more than 200 atoms. Previous calculations on medium-sized systems[72] showed that M06-2X somewhat underestimates π-π interactions. Here we find that it also generally underestimates other types of interactions in quite large systems. This may arise from the fact that M06-2X is fitted to reproduce van der Waals interactions in some small systems.[72] By adding Grimme's D3 correction, the MAD of M06-2X is reduced from 9.4 kcal/mol to 6.2 kcal/mol. Thus, it is still necessary to include D3 correction for M06-2X in predicting binding energies of large systems.

In summary, we have developed a simple and efficient distant pair correlation energy correction method for the CIM approach, which is much more accurate than the commonly used MME approximation. In the new scheme, correlation energies from distant orbital pairs beyond the distance threshold $\xi$, are extracted from electron correlation calculations of clusters as a correction term to the previous CIM correlation energy. The new CIM algorithm is simple but efficient. There is almost no extra computational cost compared to the previous CIM algorithm. Benchmark calculations show that with the distant pair correction, CIM can now recover more than 99.97% of the conventional correlation energy at various theory levels (with the default parameter). For five medium-sized complexes in the benchmark set, the deviations of binding energies calculated by the present CIM approach are always less than 2 kcal/mol with respect to the full system results. A test set, ExL8, containing of eight large complexes with up to 1027 atoms is proposed. With the binding energies computed by CIM-DLPNO-CCSD(T) or CIM-DLPNO-CCSD(T)∥RI-MP2 as the reference data, we have evaluated the performance of CIM-RI-MP2, CIM-SCS-RI-MP2 as well as four popular DFT methods. From the results, DFT with the $\omega$B97X-D functional and CIM-RI-MP2 show the best overall performance. The extension of the present CIM approach to periodic systems is undergoing.

**Acknowledgment**

This work was supported by the National Natural Science Foundation of China (Grant Nos. 21833002 and 21673110) and Scientific Research Startup Foundation of Hangzhou Normal University (Grant No. 4095C50220006). Y.G. acknowledges the financial support from the Qilu Young Scholar Program of Shandong University. Most of the calculations in this work were done on the computing facilities in the High-Performance Computing Center (HPCC) of Nanjing University and Collaborative Innovation Center of Advanced Microstructures.

**Supporting Information Available:**

Details about basis set extrapolation, benchmark results, and full geometry information are provided. The cartesian coordinates of the 13 complexes are provided in independent XYZ files.